\documentclass{aip-cp}

\usepackage[numbers]{natbib}
\usepackage{rotating}
\usepackage{graphicx}
\begin{document}

% Title portion
\title{First Study of Combined Blazar Light Curves with FACT and HAWC}

\author[aff1]{Daniela Dorner\corref{cor1}}
\author[aff2]{Robert J. Lauer\corref{cor2}}
\author[aff3]{for the FACT Collaboration: Jan Adam}
\author[aff4]{Max Ahnen}
\author[aff3]{Dominik Baack}
\author[aff5]{Matteo Balbo}
\author[aff1]{Matthias Bergmann}
\author[aff4]{Adrian Biland}
\author[aff1]{Michael Blank}
\author[aff4,aff6]{Thomas Bretz}
\author[aff3]{Kai Br\"ugge}
\author[aff3]{Jens Buss}
\author[aff5]{Anton Dmytiiev}
\author[aff3]{Sabrina Einecke}
\author[aff1]{Christina Hempfling}
\author[aff4]{Dorothee Hildebrand}
\author[aff4]{Gareth Hughes}
\author[aff3]{Lena Linhoff}
\author[aff1]{Karl Mannheim}
\author[aff4]{Sebastian M\"uller}
\author[aff4]{Dominik Neise}
\author[aff5]{Andrii Neronov}
\author[aff3]{Maximilian Noethe}
\author[aff1]{Aleksander Paravac}
\author[aff4]{Felicitas Pauss}
\author[aff3]{Wolfgang Rhode}
\author[aff1]{Amit Shukla}
\author[aff3]{Fabian Temme}
\author[aff4]{Julia Thaele}
\author[aff5]{Roland Walter}
%\eaddress[url]{http://www.aip.org}
\author{ and for the HAWC Collaboration\corref{cor4}}

\affil[aff1]{Universit\"at W\"urzburg, Institute for Theoretical Physics and Astrophysics,
Emil-Fischer-Str.\ 31, 97074 W\"urzburg,Germany}
\affil[aff2]{University of New Mexico, Department of Physics and Astronomy, 
Albuquerque, NM, USA}
\affil[aff3]{TU Dortmund, Experimental Physics 5,
Otto-Hahn-Str.\ 4, 44221 Dortmund, Germany}
\affil[aff4]{ETH Zurich, Institute for Particle Physics,
Otto-Stern-Weg 5, 8093 Zurich, Switzerland}
\affil[aff5]{University of Geneva, ISDC Data Center for Astrophysics,
Chemin d'Ecogia 16, 1290 Versoix, Switzerland}
\affil[aff6]{also at RWTH Aachen}
\corresp[cor1]{Corresponding author: dorner@astro.uni-wuerzburg.de}
\corresp[cor2]{Corresponding author: rjlauer@unm.edu}
%\corresp[cor3]{URL: \texttt{http://www.fact-project.org}}
\corresp[cor4]{For a complete author list, see 
\texttt{http://hawc-observatory.org/collaboration}}

\maketitle

\begin{abstract}
For studying variable sources like blazars, it is crucial to achieve unbiased 
monitoring, either with dedicated telescopes in pointing mode or 
survey instruments. At TeV energies, the High Altitude Water Cherenkov (HAWC) 
observatory monitors approximately two thirds of the sky every day. It uses the 
water Cherenkov technique, which provides an excellent duty cycle independent 
of weather and season. The First G-APD Cherenkov Telescope (FACT) monitors a 
small sample of sources with better sensitivity, using the imaging air 
Cherenkov technique. Thanks to its camera with silicon-based photosensors, FACT 
features an excellent detector performance and stability and extends its 
observations to times with strong moonlight, increasing the duty cycle compared 
to other imaging air Cherenkov telescopes. As FACT and HAWC have overlapping 
energy ranges, a joint study can exploit the longer daily coverage given that 
the observatories' locations are offset by 5.3 hours. Furthermore, the better 
sensitivity of FACT adds a finer resolution of features on hour-long time 
scales, while the continuous duty cycle of HAWC ensures evenly sampled 
long-term coverage. Thus, the two instruments complement each other to
provide a more complete picture of blazar variability. In this presentation, 
the first joint study of light curves from the two instruments will be shown, 
correlating long-term measurements with daily sampling between air and water 
Cherenkov telescopes. The presented results focus on the study of the 
variability of the bright blazars Mrk 421 and Mrk 501 during the
last two years featuring various flaring activities.

\end{abstract}

\section{INTRODUCTION}

\subsection{Unbiased Blazar Monitoring}

Blazars, Active Galactic Nuclei with jets oriented towards Earth, form the most 
populous class of extra-galactic sources with very high energy ($>100$~GeV) 
emission. Their very high energy (VHE) gamma-ray fluxes generally vary strongly with time scales as 
short as minutes, see for example~\cite{Aharonian2007PKS2155,Albert2007Mrk501}.
Since neither the dominating particle populations nor the location of the 
VHE accelerating regions inside these sources are known, it is crucial to 
characterize the timing parameters of varying fluxes as a clue to size and 
composition of such zones. In the VHE band, most data on blazars stem from 
Imaging Air Cherenkov Telescopes (IACT) that are often biased by observing in 
response to external alerts and are severely limited through the small field 
of view, competing targets for observation and constraints of observing in dark 
nights. Here, we include only unbiased light curve data that were 
obtained on the one hand through a fixed observation scheme with a focus on a few sources with 
FACT and on the other hand through the regular, wide field-of-view monitoring independent of environmental factors achieved by 
HAWC.

\subsection{FACT}

The First G-APD Cherenkov telescope is located at 2200\,m a.s.l.\ on the Canary Island La Palma, Spain. It is the first telescope using silicon based photosensors (SiPM, a.k.a.\ Geiger-mode Avalanche Photo-Diodes (G-APDs)) in regular operation \cite{2013JInst...8P6008A}. With a reflective area of 9.5\,sqm and the excellent photon detection efficiency of SiPMs, a trigger threshold of 350\,GeV is reached. The current analysis provides an analysis threshold of 750\,GeV \cite{2015ICRCspectrum}. 

For the long-term monitoring, a camera with SiPMs is ideally suited for several reasons. Being robust and stable, SiPMs provide a stable detector performance \cite{2014JInst...9P0012B} facilitating the automation of operation which leads to high data-taking efficiency increasing the duty-cycle of the instrument. Also the fact that SiPMs do not age when exposed to bright light helps increasing the duty-cycle as observations during bright moon light are carried out in standard operation mode \cite{factmoon}. This not only enlarges the total amount of observation time but also minimizes the gaps around full-moon providing a more continuous data sample \cite{2015ICRCmonitoring}. 

\subsection{HAWC}

The HAWC Observatory is located at an elevation of 4,100~m above sea level on the slope of the Sierra Negra volcano in the state of Puebla, Mexico (18$^\circ$59'41"N 97$^\circ$18'30.6"W). It has been operating since March 2015 in its completed configuration consisting of 300 Water Cherenkov Detectors (WCDs) that instrument an area of 22,000~m$^2$. The array is sensitive to extensive air showers induced by gamma rays with energies between approximately 100~GeV and 100~TeV with a peak sensitivity in the range 2 to 10 TeV, depending on source declinations and spectra. HAWC's main distinctions when compared to other currently operating VHE gamma-ray detectors is its near continuous duty cycle $\sim90$~\% and its wide field of view of $\sim2$~steradians. HAWC can monitor any source that transits through a $45^{\circ}$ cone centered on local zenith for up to 6 hours per day. For more details on the design, operation, and data processing see \cite{HawcSensitivity,design-icrc2015}.

\section{LIGHT CURVE CORRELATION}

\subsection{Data Sample}

Having an offset of 5.3 hours between the two sites, the data samples of FACT 
and HAWC provide the possibility to combine the light curves extending the 
continuous observation time to up to 12 hours. 
%(can we say this like that? this
%number is based on the +-45deg zd) 
While FACT has a better sensitivity and therefore timing resolution, the HAWC 
monitoring is continuous independent of weather conditions at the site. 
As a basis for this first study, a time range of a bit more than one year 
(from November 26, 2015 until December 9, 2015, i.e.\ 378 nights) 
was chosen based on the available HAWC data sample. 

The study is focusing on three sources: the Crab Nebula, which is ideal for 
comparing the two instruments as at TeV energies no flux variations have been
detected so far, and the two bright TeV blazars Mrk\,421 and Mrk\,501. 

\subsection{FACT Analysis}

In the meantime, 4-5 years of monitoring data from FACT are available. In total more than 7400~hours of data have been collected so far where Mrk\,421 has been observed for more than 1400~hours and Mrk\,501 for about 1800~hours (status 27.9.2016).  For the study presented here, the time range was limited to the dates determined by the first processing of HAWC data. During 
that time, FACT observed the Crab Nebula in 97 nights for a total of 220 hours, 
Mrk\,421 in 136 nights for a total of 315 hours and Mrk\,501 in 177 nights for a 
total of 441 hours. 

%FROM DB
%all
% crab 97, 219.9h
% 421: 136, 315.4h -> 2.3 h/night
% 501: 177, 440.6h -> 2.49 h/night
%after zd/th cuts:
% crab: 88, 175.7h (72 with tighter cuts, 118.3h) 
% 421: 120, 257.2h
% 501: 145, 325.4h
%after all cuts:
% crab: 81, 147.6h (65 with tighter cuts, 97.2h)
% 421: 113, 234.0h -> 2.1h/night
% 501: 136, 274.0h -> 2h/night
%only nights with > 20 min of data (LC plots)
% crab: 74, (57, )
% 421: 103, 
% 501: 130, 
%FROM macros:
%only nights with > 20 min of data (LC plots)
% 421: 101, 231.8h -> 2.3 h/night
% 501: 125, 270.6h -> 2.16h/night
%common nights: (correlation plot) 
% crab: 
% 421: 88, 213.0h -> 2.42 h/night
% 501: 120, 251.5h -> 2.09 h/night
%crab correlation study 
% crab: 54, 91.0h -> 1.68 h/night

% hawc: lc: 
% 421: 327
% 501: 333

Based on data of the Crab Nebula, which features a stable flux at TeV energies, 
the dependency of the background corrected count rates on observational 
parameters like zenith distance and trigger threshold were studied. Based on this, 
the analyzed data were restricted to a range of zenith distance and trigger 
threshold in which the count rate of the Crab Nebula is stable. 
Using the count rate determined for the Crab Nebula, the count rates for Mrk\,421 
and Mrk\,501 were converted to Crab Units. For this, time ranges with different
detector or analysis setup were treated independently. 

The analysis was performed as described in \cite{2015arXiv150202582D}, and bad quality data were rejected with a data selection algorithm as described in \cite{2013arXiv1311.0478D}. 
%also these two references are proceedings 
Selecting data with good quality, small zenith distance and small trigger threshold, the study included for Mrk\,421 a total of 232~hours from 101 nights, i.e.\ in average 2.3 hours per night, and for Mrk\,501 a total of 271~hours of data from 125 nights, i.e.\ in average 2.2~hours per night. 

\subsection{HAWC Analysis}

The analysis of VHE flux light curves from HAWC data binned in daily intervals 
(one transit, $\sim 6$~hours) is described in~\cite{HawcMonitoringGamma}.
For this correlation analysis, we used HAWC data collected between November 26, 
2015, and December 9, 2015. 
We excluded those days from the the analysis of a 
given source for which the transit coverage was less than 75~\% due to 
interruption of data taking, primarily caused by maintenance or construction.
For the analysis of Mrk 421, 327 of 378 transits were included. The daily flux 
was calculated based on a maximum likelihood fit of the normalization for a spectrum defined by 
$ dN/dE = F_i \cdot ( E/ 1~\mbox{TeV} )^{-2.3} \cdot \exp{(-E / 5~\mbox{TeV})}$.
In the case of Mrk 501, 333 of 378 transits passed the quality cuts, and the 
spectral shape for the flux calculation was fixed to
$ dN/dE = F_i \cdot ( E/ 1~\mbox{TeV} )^{-1.7} \cdot \exp{(-E / 5~\mbox{TeV})}$.

In order to allow a comparison of absolute photon flux values to those provided 
by FACT, we performed an analytical integration of these spectra above 1~TeV and converted the resulting photon flux into Crab Units (CU) by dividing 
through the photon flux observed with HAWC for the Crab Nebula, averaged over 
the same time period as included in this analysis.

\subsection{Combined Sample}

For a correlation study, nights with data from both instruments were used, including 54 nights from the Crab Nebula, 88 nights from Mrk\,421 and 120 nights of Mrk\,501. The observation time per night for HAWC is $\sim 6$ hours for a full transit. In case of FACT for the combined data sample after all cuts, 2.4~hours per night were included in the study for Mrk\,421 and 2.1 hours per night were for Mrk\,501. 

\section{RESULTS AND DISCUSSION}

\begin{figure}
 \includegraphics[width=1.0\textwidth]{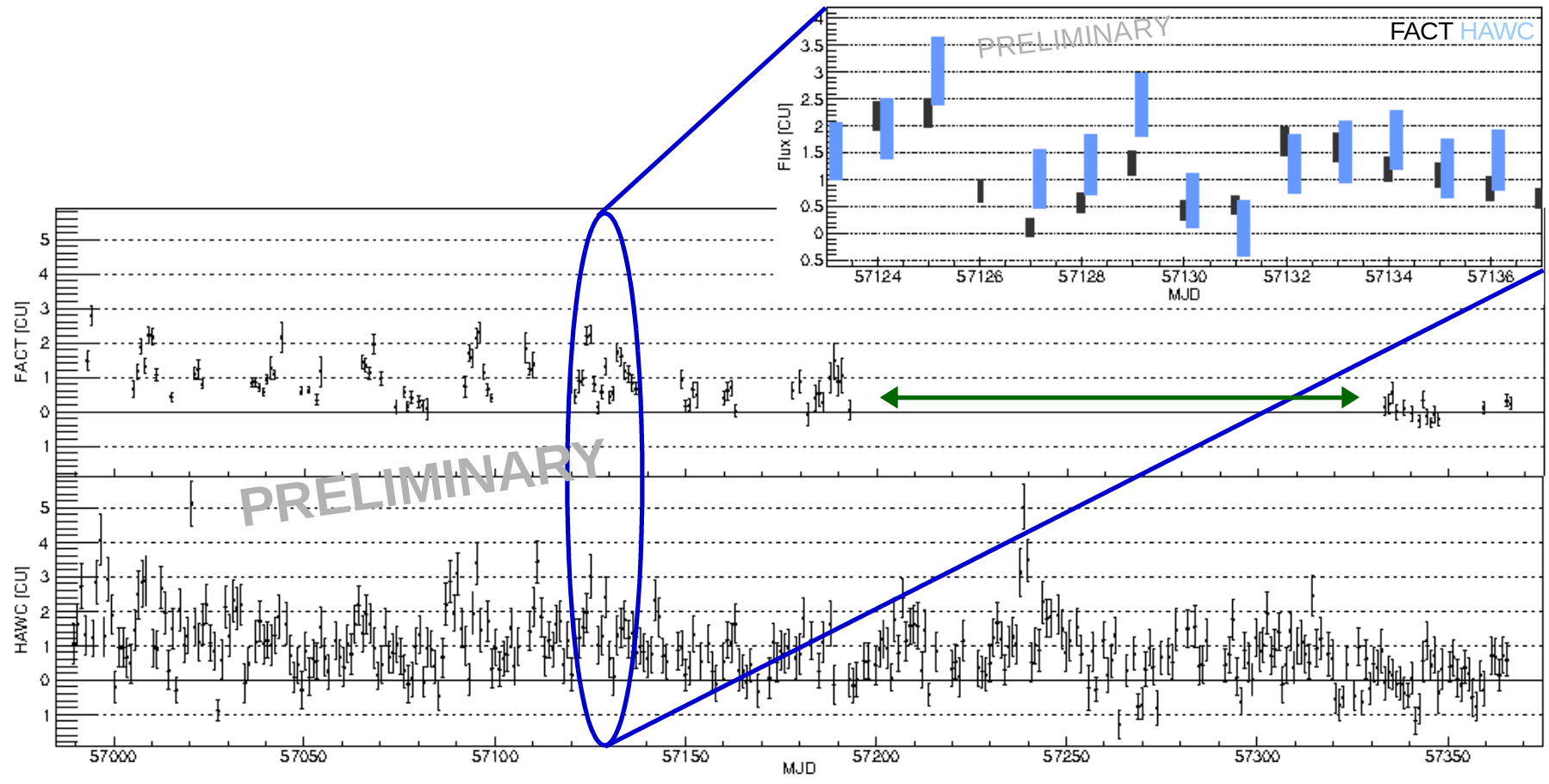}
\label{fig:mrk421}
\caption{Daily light curve for Mrk\,421 measured by FACT (top) and by HAWC (bottom) from November 26, 2014 until December 9, 2015. The inset shows a zoom-in, where in black the FACT observations are shown and in blue the HAWC observation. The width of each box in y-direction indicates the flux error and the one in x-direction the duration of the observation. The green arrow indicates the observational gap when the source is not visible during the night.}
\end{figure}

\begin{figure}
 \includegraphics[width=1.0\textwidth]{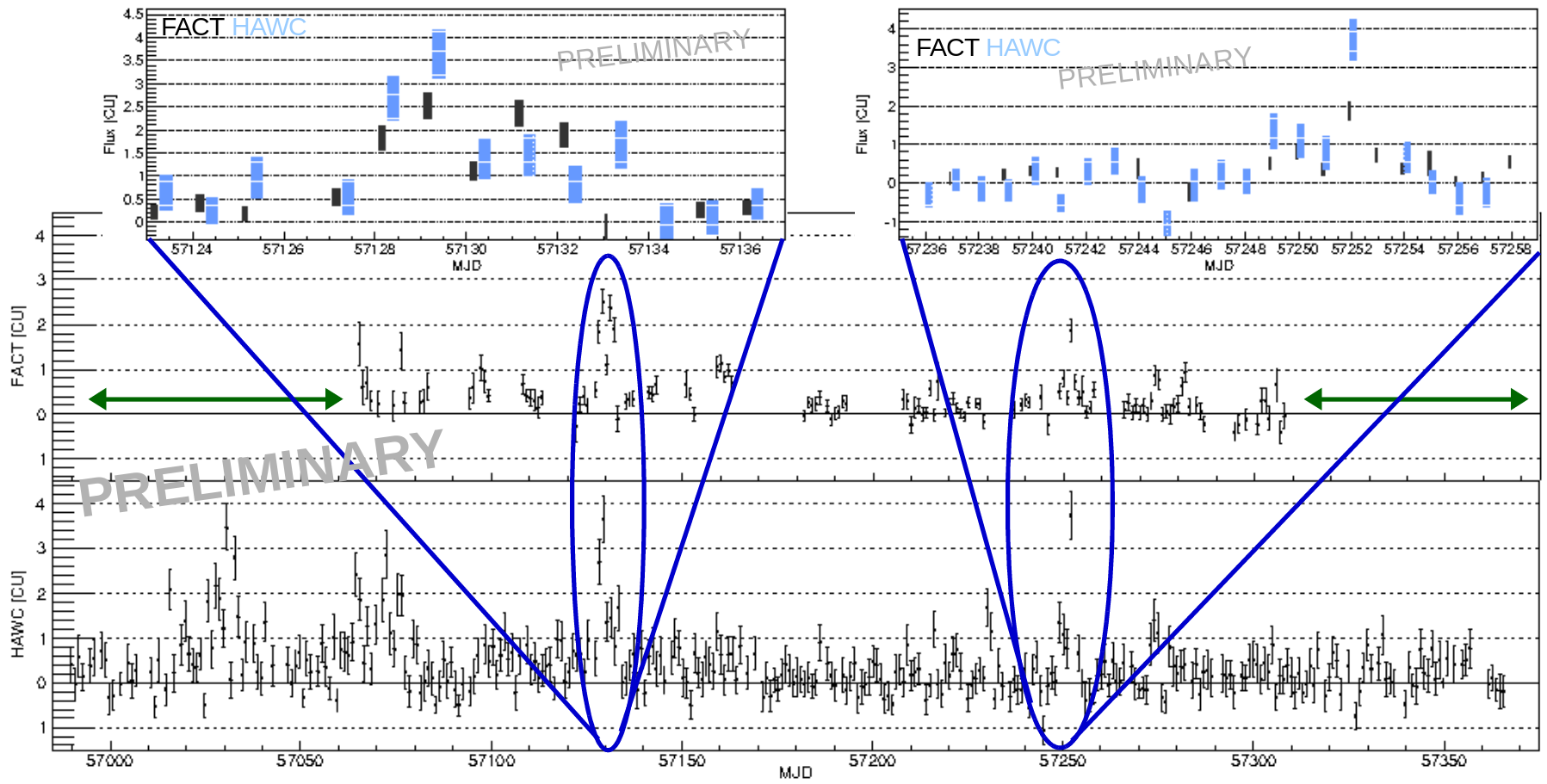}
\label{fig:mrk501}
\caption{Daily light curve for Mrk\,501 measured by FACT (top) and by HAWC (bottom) from November 26, 2014 until December 9, 2015. The insets show zoom-ins on flaring activities, where in black the FACT observations are shown and in blue the HAWC observation. The width of each box in y-direction indicates the flux error and the one in x-direction the duration of the observation. The green arrows indicate the observational gaps when the source is not visible during the night.}
\end{figure}

Combining the data of the two instruments, the daily light curves for Mrk\,421 and Mrk\,501 have been compiled as shown in Figures \ref{fig:mrk421} and \ref{fig:mrk501}. We can clearly see the complementary features of HAWC (bottom panels) and FACT (top panels) in the direct comparisons of the light curves. In the insets, zooms to smaller time ranges are shown where FACT data are plotted in black and HAWC data in blue, and the box-width corresponds to the duration of the observation. The statistical errors for FACT data are generally smaller due to a lower energy threshold, better background suppression and better point spread function, while the HAWC data covers the whole period continuously and does not have seasonal gaps. Additional gaps for FACT around full moon are due to safety requirements. While for HAWC the observation time per transit is 6 hours, the duration of the FACT observation ranges from 40 minutes at the beginning and end of the observing season to 8 hours in the middle. The average observation night is 2.4~hours for Mrk\,421 and 2.1~hours for Mrk\,501 for the results presented here. 
The insets highlight the extension of nightly observation window from $\leq6$ hours with FACT or HAWC alone to up to 12 hours with the combined data. 

%The zoomed insets of periods of increased source activity show a direct comparison of the CU flux values for selected time ranges from both experiments. They highlight the extension of nightly observation window from $\leq6$ hours with FACT or HAWC alone to up to 12 hours with the combined data. Also the different exposures of the instruments are visible. 

The analytical conversion of fluxes to Crab units for HAWC data allows an approximate comparison with the FACT excess counts converted to CU, but a systematic uncertainty remains for the absolute scaling because of the non-uniform definition of these flux units.
The flux in CU depends both on the spectral assumptions for the blazar and Crab spectra in the HAWC analysis and the relative thresholds in HAWC and FACT for defining an integrated photon flux. Also spectral variations are not considered so far. We are currently working towards a detailed comparison of the spectral shapes observed in both experiments that will allow us to constrain this uncertainty in the absolute flux scaling.

%%correlation plot comments removed:
%- for correlation plots, selection of nights (i.e. exclusion of HAWC day time 
%data) where both instruments have data (88 for Mrk 421) 
%- deviations from strong correlation possibly due to different thresholds; 
%varying spectra and hence different response; variations on time scales of 
%hours or less

%%correlation plot comments removed:
%- for correlation plots, selction of nights (i.e. exclusion of HAWC day time 
%data) where both instruments have data (120 for Mrk 501) 

\section{SUMMARY AND OUTLOOK}
 
To assess the characteristics of blazar flares unbiased monitoring is crucial. The long-term monitoring with FACT and HAWC provide large data sample, of which in this study one year of data was analyzed for the first time together. Comparison with a daily binning shows that the results are comparable. Furthermore, the two instruments nicely complement each other where HAWC provides a more continuous data sample over the whole year and FACT a better timing resolution thanks to its better sensitivity. 
%Differences in the absolute flux can be due to different energy thresholds. 

Further investigations will include more detailed analyses and a larger data sample. For FACT, the calculation for fluxes based on Monte Carlo simulations will be included. The differences in the analyses and sensitivities will be studied using data from the Crab Nebula. For a direct comparison, also strictly simultaneous data might be used as about half an hour of observational overlap is expected for part of the nights. These studies will give a first insight of a cross-calibration between the air-Cherenkov and water-Cherenkov technique. 
Once the cross-calibration of the analyses is done, the long-term light curves will be studied for the blazars Mrk\,421 and Mrk\,501, investigating both temporal behavior and correlations, taking into account the time difference between the two sites.

% 
% Cite all figures in the text consecutively. The word ``Figure'' should be spelled out if it is the first word of the sentence and abbreviated as ``Fig.'' elsewhere in the text. Place the figures as close as possible to their first mention in the text at the top or bottom of the page with the figure caption positioned below the figure, all centered. Figures must be inserted in the text and may not follow the Reference section. Set figure captions in 9 point size, Times Roman font. Type the word ``FIGURE 1.'' in bold uppercase, followed by a period.\footnote{This is an example of a footnote.}

% Acknowledgement
\section{ACKNOWLEDGMENTS}

FACT Collaboration:
The important contributions from ETH Zurich grants ETH-10.08-2 and ETH-27.12-1 as well as the funding by the German BMBF (Verbundforschung Astro- und Astroteilchenphysik) and HAP (Helmholtz Alliance for Astroparticle Physics) are gratefully acknowledged. We are thankful for the very valuable contributions from E. Lorenz, D. Renker and G. Viertel during the early phase of the project. We thank the Instituto de Astrofisica de Canarias allowing us to operate the telescope at the Observatorio del Roque de los Muchachos in La Palma, the Max-Planck-Institut f\"r Physik for providing us with the mount of the former HEGRA CT3 telescope, and the MAGIC collaboration for their support.\\
HAWC Collaboration:
We acknowledge the support from: the US National Science Foundation (NSF); the US Department of Energy Office of High-Energy Physics; the Laboratory Directed Research and Development (LDRD) program of Los Alamos National Laboratory; Consejo Nacional de Ciencia y Tecnolog\'{\i}a (CONACyT), Mexico (grants 271051, 232656, 167281, 260378, 179588, 239762, 254964, 271737); Red HAWC, Mexico; DGAPA-UNAM (grants RG100414, IN111315, IN111716-3, IA102715, 109916); VIEP-BUAP; the University of Wisconsin Alumni Research Foundation; the Institute of Geophysics, Planetary Physics, and Signatures at Los Alamos National Laboratory; Polish Science Centre grant DEC-2014/13/B/ST9/945. 

% References

%\nocite{*}
\bibliographystyle{aipnum-cp}%
\bibliography{fact-hawc_gamma2016}%

\end{document}